\documentclass[envcountsame,envcountsect,orivec,12pt,undated,nolinenumbers]{lnthi}
\usepackage{amsmath,graphics}

\title{Still Simpler Static Level Ancestors}

\author{Torben Hagerup}

\institute{\Tinfuna[5]\\
  \email{hagerup@informatik.uni-augsburg.de}}

\def\Tvn#1{\hbox{\textit{#1\/}}} 
\def\Tvnn#1{\hbox{\textit{\scriptsize #1\/}}} 
\def\Tfloor#1{\lfloor #1\rfloor}
\def\Tceil#1{\lceil #1\rceil}
\def\Tffloor#1{\lfloor\mskip-4.5mu\lfloor#1\rfloor\mskip-4.5mu\rfloor}
\def\Tsub#1{_{\mbox{\scriptsize #1}}} 
\def\Ttwodots{\mathinner{\ldotp\ldotp}}
\def\hax{\widehat{x}}
\def\haxx{\rlap{\lower1.5pt\hbox{$\widehat{\phantom{x}}$}}{x}}
\def\tixx{\rlap{\lower1.5pt\hbox{$\widetilde{\phantom{x}}$}}{x}}

\begin{document}
  \overfullrule=5pt  

\maketitle{}

\begin{abstract}
A level-ancestor or LA query about a rooted tree $T$
takes as arguments a node $v$ in $T$, of depth $d_v$,
say, and an integer $d$ with $0\le d\le d_v$ and
returns the ancestor of $v$ in $T$ of depth~$d$.
The static LA problem is to process
a given rooted tree $T$ so as to support efficient
subsequent processing of LA queries about~$T$.
All previous efficient solutions to the static
LA problem work by reducing a given instance of
the problem to a smaller instance of
the same or a related problem,
solved with a less efficient data structure, and
a collection of small micro-instances for
which a different solution is provided.
We indicate the first efficient solution to the static
LA problem that works directly, without
resorting to reductions or
micro-instances.


\end{abstract}

\section{Introduction}
\label{sec:intro}%

A \emph{level-ancestor}
or \emph{LA query} about a rooted tree $T$ takes
as arguments
a node $v$ in $T$, of depth $d_v$, say,
and an integer $d$ with $0\le d\le d_v$
and returns the ancestor of $v$ in $T$
of depth $d$
(or, in some formulations,
of depth $d_v-d$).
LA queries have applications, e.g., to the computation
of semigroup sums over paths in trees~\cite{Cha87},
the aggregation of minima over subtrees~\cite{AtaY09}
and the recognition of breadth-first-search
trees~\cite{HagN85,Man90}.
They are also considered part of the repertoire
of operations that a well-endowed data structure
for representing rooted trees should support
(see, e.g.,~\cite{Nav16}).

One distinguishes between static and dynamic versions
of the \emph{LA problem} of supporting efficient LA queries.
In dynamic versions of the
problem~\cite{AlsH00,Die91,NavS14}, LA queries are interspersed
with calls of operations that change the structure
of the underlying tree.
In the static version of the problem,
which forms the focus of the present text, the tree
is given once and for all, and the task is to preprocess
it so that subsequent LA queries can be executed fast.

The static LA problem was first
considered by Berkman and Vishkin~\cite{BerV94},
who reduced it
to a problem about a sequence of integers.
Specifically, suppose that a depth-first search
(DFS) of a rooted tree $T$ appends
(the name of) a node $v$ in
$T$ to an initially empty sequence
whenever $v$ is visited for the first time or the DFS
withdraws to~$v$.
This yields a sequence $P_1$ of $2 n-1$ nodes.
The DFS can also mark each node $v$ with its depth
in $T$ and with a position in $P_1$
in which $v$ occurs.
Replacing each node in $P_1$ by its depth in~$T$
yields a new sequence $P_2$ of integers.
The ancestor in depth $d$ of a node $v$ in $T$
of depth $d_v$, where $0\le d\le d_v$, can now
be found as the node in $P_1$ in the same
position as the first occurrence in $P_2$
of a number bounded by $d$ in or following the
position that in $P_1$ contains an arbitrary
occurrence of~$v$.
To answer LA queries about $T$, it therefore
suffices to be able to answer FS
(``find-smaller'') queries
about $P_2$, where an \emph{FS query} about
a sequence $P=(d_0,\ldots,d_{m-1})$
of integers takes
as arguments integers $i$ and $d$ and returns
$\min(\{j\in\{0,\ldots,m-1\}\mid j\ge i$ and $d_j\le d\}\cup\{m\})$.

Say that a
sequence $Y$ is a \emph{1-difference sequence}
if $Y=(y_0,\ldots,y_{n-1})$ for integers
$y_0,\ldots,y_{n-1}$ 
with the property that $|y_i-y_{i-1}|\le 1$
for $i=1,\ldots,n-1$.
Of course, $P_2$ above has this property.
Berkman and Vishkin gave a family of
parallel algorithms that
input a 1-difference sequence $Y$
of length $n$
and output a data structure of
$O(n)$ words
that enables subsequent FS queries about $Y$ to
be answered in constant time by a single processor.
Here and in the following, when the space
requirements of a data structure are expressed in
terms of words, we use the common convention
that a word consists of $\Theta(\log n)$ bits.
For $k\ge 1$, the $k$th algorithm 
in the family works in
$O(\log^{(k)} n)$ time using
$O({n/{\log^{(k)}n}})$ processors,
where $\log^{(k)}$ denotes the $k$-fold
iterated logarithm function.
A central idea is to equip each
$i\in\{1,\ldots,n-1\}$ with precomputed
answers to certain FS queries with
first argument~$i$, choosing
the number of such precomputed answers to
be proportional to the highest power of~2
that divides~$i$.
The result of Berkman and Vishkin
implies the existence of a sequential
algorithm to carry out the preprocessing
for FS queries in $O(n)$ time.
Such an algorithm, without the complications
necessary in a parallel setting, was described
by Ben-Amram~\cite{Ben09}.

A different approach to the LA problem was
initiated by Dietz~\cite{Die91}.
Here a main idea is to decompose the given
tree into a set of paths, to provide complete
ancestor information within each path in an
array and to introduce a mechanism that allows
an LA query to find its relevant path
(the one that contains the node to be returned)
in constant time.
A simpler data structure based on the same idea
was described by Bender
and Farach-Colton~\cite{BenF04}.
Yet another data structure, intermediate
in complexity, was proposed by
Alstrup and Holm~\cite{AlsH00}.
All three data structures occupy $O(n)$ words
and can be constructed in $O(n)$ time, just as the
data structure of Ben-Amram.

All of the data structures discussed above have
at their core a less efficient data structure,
and they work by reducing a given instance of the LA or
FS problem
to a smaller instance of the same or a related problem, which
is handled with the less efficient data structure,
and a collection of small
\emph{micro-instances}, for which a different
solution is provided.
More concretely, in the case of the sequential
solutions~\cite{AlsH00,Ben09,BenF04,Die91},
what we call
the \emph{basic} data structure has a logarithmic overhead
and needs $\Theta(n\log n)$ preprocessing time and
$\Theta(n\log n)$ words of space to handle input
instances of size~$n$.
The tree-based solutions~\cite{AlsH00,BenF04,Die91}
partition the given tree into a collection
of \emph{micro-trees} of $O(\log n)$ nodes
each, ``held together'' by
a \emph{macro-tree} with $O({n/{\log n}})$ nodes,
the macro-tree is stored in an instance of
the basic data structure, the micro-trees are
handled with table lookup,
and it is shown how to process a top-level query
with a constant number of queries in the macro-tree
and in micro-trees.
In the data structure of
Ben-Amram~\cite{Ben09}, the separation between
the original instance and a macro-instance is
less clear-cut, but there are still
micro-instances of size $O(\log n)$
handled with table lookup.

We describe a new data structure
for the static LA problem
that works directly, without resorting to
reductions or micro-instances,
and is the first solution to the LA problem
with this property.
In order to highlight what sets the new structure
off from its predecessors, we call it
the \emph{one-level} structure.
Like the data structures of
Berkman and Vishkin~\cite{BerV94} and Ben-Amram~\cite{Ben09},
the one-level structure actually solves the more
general FS problem for 1-difference sequences.
While our result does not allow us to prove any
new asymptotic bounds, we expect
the one-level structure to be easier to program
and to perform better in practice than the
known data structures with the same
guaranteed resource bounds.
The paper by Bender and Farach-Colton~\cite{BenF04}
has been cited more than a hundred times,
according to Google Scholar.
It seems likely that most or all of the applications of
solutions to the LA problem described in the
scientific literature can benefit from the results
developed here.

In fact, we prefer to phrase the discussion in terms
of \emph{FL} (``find-larger'') queries defined in
complete analogy with FS queries, i.e.,
an FL query about a sequence $Y=(y_0,\ldots,y_{n-1})$
of integers inputs integers $x$ and $y$ and returns 
$\Tvn{FL}(x,y)=
\min(\{i\in\{0,\ldots,n-1\}\mid i\ge x$ and $y_i\ge y\}\cup\{\bot\})$,
where $\bot$ is a default
value, considered larger than $n-1$,
to be returned when the query has
no natural answer.
Correspondingly, we speak of the one-level FL structure.
We believe that the choice
of FL over FS leads to more natural
intuition and terminology, human beings generally
having more experience being above
hill sides than being below cave sides.
Informally, let us associate with a sequence
$(y_0,\ldots,y_{n-1})$ of integers the sequence
$((0,y_0),\ldots,(n-1,y_{n-1}))$ of points in the
Euclidean plane and imagine these connected with
the line segments $((i-1,y_{i-1}),(i,y_i))$,
for $i=1,\ldots,n-1$, to create the contour of a landscape.
If an FL query with arguments $x$ and $y$,
which we will write simply as $(x,y)$,
is \emph{nontrivial}, i.e., if
$0\le x<n$ and $y_x<y\le\max_{0\le i<n}y_i$, then it
can be answered by reporting the x coordinate
of the point in the landscape visible by looking
horizontally to the right from the point $(x,y)$
($\bot$ if there is no such point).

\section{The One-Level FL Structure}

The new one-level FL structure combines ideas of the
basic data structures of Ben-Amram~\cite{Ben09} and
Bender and Farach-Colton~\cite{BenF04}, even though
this may not be apparent at a first inspection.
It operates with the notion of \emph{valleys}.
Given a sequence $(y_0,\ldots,y_{n-1})$ of $n$
arbitrary integers and a pair $(x,y)$ of integers
with $0\le x<n$ and $y\ge y_x$, informally,
the valley of $(x,y)$ is the x coordinate of the
rightmost deepest point that one can reach from
$(x,y)$ while moving only downwards and to the left
and staying above the contour of the landscape.
Formally, say that a point $(\overline{x},\overline{y})$
is \emph{down-left reachable} from $(x,y)$
if $\overline{x}$ and $\overline{y}$ are integers
such that $0\le\overline{x}\le x$, $y_{\overline{x}}\le\overline{y}\le y$,
and $y_i<y$ for all integers $i$ with $\overline{x}\le i<x$.
Of course, $(x,y)$ is down-left reachable from itself.
To define the valley of $(x,y)$, where
$x$ and $y$ are integers with $0\le x<n$ and $y\ge y_x$,
let $\overline{y}$ be minimal such that some point of
the form $(\overline{x},\overline{y})$ is down-left
reachable from $(x,y)$.
Then the valley of $(x,y)$ is the largest
$\overline{x}\in\{0,\ldots,x\}$ with $y_{\overline{x}}=\overline{y}$.

\subsection{Initialization}

When initialized with a 1-difference sequence
$Y=(y_0,\ldots,y_{n-1})$,
the one-level FL structure first computes an array
$\Tvn{Valley}[0\Ttwodots n]$ such that $\Tvn{Valley}[x]$
is the valley of $(x,y_x)$ for all
$x\in\{0,\ldots,n-1\}$ and $\Tvn{Valley}[n]$
has the artificial value $n-1$.
Even if $Y$
is a sequence of $n$ arbitrary integers
(or real numbers, given a suitably
generalized definition of valleys),
this can be done in $O(n)$ time
with a sweep over $0,\ldots,n-1$ shown in
Fig.~\ref{code:valley}.
When the sweep is at some $x\in\{0,\ldots,n-1\}$,
its state is given by the sorted sequence of all valleys
of points of the form $(x,y)$, where $y\ge y_x$,
with each valley $\overline{x}$ represented by the
triple $(\overline{x},y_{\overline{x}},
\max\{y_i\mid\overline{x}\le i\le x\})$.
The triples, whose components, in the order from
left to right, are referred to using the field
names $x$, \Tvn{low} and \Tvn{high}
in the code, are stored
in order in an array $S$, preceded by the
dummy triple $(0,-\infty,\infty)$.
$S$ is manipulated as a stack, except that the
sweep occasionally inspects the \Tvn{high} component
of the triple just below the top triple.
The computation of valleys corresponds roughly
to the decomposition of the given tree
into paths in the algorithm of
Bender and Farach-Colton~\cite{BenF04}.

\begin{figure}
\begin{tabbing}
\quad\=\quad\=\quad\=\quad\=\kill
\>$\Tvn{top}:=0$; $(*$ stack pointer of $S$ $*)$\\
\>$S[\Tvn{top}]:=(0,-\infty,\infty)$;
 $(*$ dummy sentinel; will never be popped $*)$\\
\>\textbf{for} $x:=0$ \textbf{to} $n-1$ \textbf{do}
 $(*$ sweep from left to right $*)$\\
\>\>\textbf{while} $S[\Tvn{top}].\Tvn{low}\ge y_x$
 \textbf{do} $\Tvn{top}:=\Tvn{top}-1$;
 $(*$ pop valleys no deeper $*)$\\
\>\>\textbf{if} $S[\Tvn{top}].\Tvn{high}\ge y_x$ \textbf{then}
 $(*$ cannot reach even first valley from $(x,y_x)$ $*)$\\
\>\>\>$\Tvn{Valley}[x]:=x$; $(*$ cannot go deeper from $(x,y_x)$ $*)$\\
\>\>\>\textbf{if} $S[\Tvn{top}].\Tvn{high}>y_x$ \textbf{then}\\
\>\>\>\>$\Tvn{top}:=\Tvn{top}+1$; $(*$ push $*)$\\
\>\>\>\>$S[\Tvn{top}]:=(x,y_x,y_x)$; $(*$ $x$ is a new valley $*)$\\
\>\>\textbf{else}
 $(*$ a deeper valley can be reached from $(x,y_x)$ $*)$\\
\>\>\>\textbf{while} $S[\Tvn{top}-1].\Tvn{high}<y_x$
 \textbf{do} $\Tvn{top}:=\Tvn{top}-1$;
 $(*$ pop until before $\ge$ $*)$\\
\>\>\>$\Tvn{Valley}[x]:=S[\Tvn{top}].x$;
 $(*$ deepest reachable valley $*)$\\
\>\>\>\textbf{if} $S[\Tvn{top}-1].\Tvn{high}>y_x$
 \textbf{then} $S[\Tvn{top}].\Tvn{high}:=y_x$;
 $(*$ have to pass here $*)$\\
\>\>\>\textbf{else} $\Tvn{top}:=\Tvn{top}-1$;
 $(*$ top valley not deepest once sweep continues $*)$\\
\>$\Tvn{Valley}[n]:=n-1$;
 $(*$ special convention $*)$
\end{tabbing}
\caption{The computation of the array $\Tvn{Valley}$
 for a sequence $(y_0,\ldots,y_{n-1})$ of integers.}
\label{code:valley}
\end{figure}

For every integer $x\ge 1$, let $\pi(x)$ be
the largest power of~2 that divides~$x$.
The idea of using $\pi$ in the solution of
the FS or FL problem goes back to
Berkman and Vishkin~\cite{BerV94} and
Ben-Amram~\cite{Ben09}, but it is crucial
to our approach to use $\pi$ in a different way.
The one-level FL structure is parameterized by an
integer constant $\kappa\ge 3$.
Its initialization proceeds to compute two arrays
$\Tvn{Weight}[0\Ttwodots n-1]$ and
$\Tvn{Jump}[0\Ttwodots n-1]$ such that
$\Tvn{Weight}[\overline{x}]=
|\{x\in\{\overline{x},\ldots,n-1\}:\Tvn{Valley}[x]=\overline{x}\}|$
for all $\overline{x}\in\{0,\ldots,n-1\}$
and
$\Tvn{Jump}[\hax]=\Tvn{Valley}
 [\Tvn{FL}(\hax,y_{\haxx}+(\kappa-2)\pi(\hax))]$
for all $\hax\in\{1,\ldots,n-1\}$,
while $\Tvn{Jump}[0]$ is set to the artificial
value~0.
Following Bender and Farach-Colton~\cite{BenF04},
we define a \emph{ladder} of \emph{height} $h$
\emph{located at} $x$, where $h$ and $x$ are
integers with $h\ge 0$ and $0\le x<n$,
to be an array with index set $\{y_x+1,\ldots,y_x+h\}$
that maps each $y\in\{y_x+1,\ldots,y_x+h\}$ to
$\Tvn{FL}(x,y)$.
The initialization of the structure is
finished by equipping each $x\in\{1,\ldots,n-2\}$
with a ladder $L_x$ located at $x$ and of height
$\min\{\max\{\kappa-1,\kappa'(\Tvn{Weight}[x]-1)-2\},
 y\Tsub{max}-y_x\}$,
where $\kappa'=\Tceil{{{(2\kappa+2)}/{(\kappa-2)}}}$
and $y\Tsub{max}=\max\{y_0,\ldots,y_{n-1}\}$,
and each $x\in\{0,n-1\}$ with a ladder
$L_x$ of height $y\Tsub{max}-y_x$.
This is easy to do in a second sweep over $0,\ldots,n-1$,
this time from right to left.
The complete initialization of the one-level FL structure
for a 1-difference sequence $(y_0,\ldots,y_{n-1})$
is shown in Fig.~\ref{code:basic}, which assumes
that the default value $\bot$ is chosen as~$n$.

\begin{figure}
\begin{tabbing}
\quad\=\quad\=\quad\=\quad\=\kill
\>Compute $\Tvn{Valley}[0\Ttwodots n]$;
 $(*$ as in Fig.~\ref{code:valley} $*)$\\
\>$\Tvn{Weight}[0\Ttwodots n-1]:=[0,\ldots,0]$;
$(*$ initialize counts to zero $*)$\\
\>\textbf{for} $x\in\{0,\ldots,n-1\}$ \textbf{do}
 $\Tvn{Weight}[\Tvn{Valley}[x]]:=\Tvn{Weight}[\Tvn{Valley}[x]]+1$;\\
\>$y\Tsub{min}=\min\{y_0,\ldots,y_{n-1}\}$;
  $y\Tsub{max}=\max\{y_0,\ldots,y_{n-1}\}$;\\
\>$\Tvn{RightSight}[y\Tsub{min}\Ttwodots y\Tsub{max}+1]
 :=[n,\ldots,n]$;
 $(*$ initial default of sweepline ($\bot=n$) $*)$\\
\>\textbf{for} $x:=n-1$ \textbf{downto} 0 \textbf{do}
 $(*$ sweep from right to left $*)$\\
\>\>$\Tvn{RightSight}[y_x]:=x$;
 $(*$ update sweepline $*)$\\
\>\>$h:=y\Tsub{max}-y_x$;
 $(*$ upper bound on ladder height; tight for $x\in\{0,n-1\}$ $*)$\\
\>\>\textbf{if} $0<x<n-1$ \textbf{then}
$h:=\min\{\max\{\kappa-1,\kappa'(\Tvn{Weight}[x]-1)-2\},h\}$;\\
\>\>$L_x[y_x+1\Ttwodots y_x+h]:=\Tvn{RightSight}[y_x+1\Ttwodots y_x+h]$;
 $(*$ construct ladder at $x$ $*)$\\
\>\>\textbf{if} $x=0$ \textbf{then} $\Tvn{Jump}[x]:=0$;
 $(*$ special convention $*)$\\
\>\>\textbf{else}
$\Tvn{Jump}[x]:=\Tvn{Valley}[\Tvn{RightSight}[\min\{y_x+
 (\kappa-2)\pi(x),y\Tsub{max}+1\}]]$;
\end{tabbing}
\caption{The initialization of the one-level FL structure
 for a 1-difference sequence $(y_0,\ldots,y_{n-1})$.}
\label{code:basic}
\end{figure}

\subsection{Processing of Queries}

For $x\ge 1$, let
$\Tffloor{x}=2^{\Tfloor{\log_2 x}}$, i.e.,
$\Tffloor{x}$ is the largest power of~2
no larger than~$x$.
To answer a nontrivial query $(x,y)$,
the one-level FL structure computes $t=y-y_x$ and,
if $t<\kappa$, returns $L_x[y]$,
which is obviously correct.
If $t\ge\kappa$, it returns
$L_{\Tvnn{Jump}[\mskip 1mu\haxx\mskip 3mu]}[y]$,
where $\hax$ is the largest integer
with $1\le\hax\le x$ and $\pi(\hax)=\Tffloor{{t/\kappa}}$
if there is such an integer, and
$\hax=0$ if not.
This procedure, augmented with instructions
to handle trivial queries,
is shown in Fig.~\ref{code:query}.

\begin{figure}
\begin{tabbing}
\quad\=\quad\=\quad\=\quad\=\kill
\>\textbf{if} $x\ge n$ or $y>y\Tsub{max}$
 \textbf{then return} $n$;
 $(*$ default value $\bot$ $*)$\\
\>\textbf{if} $x<0$ \textbf{then} $x:=0$;
 $(*$ no effect on result $*)$\\
\>\textbf{if} $y\le y_x$ \textbf{then return} $x$;
 $(*$ trivial case $*)$\\
\>\textbf{if} $y<y_x+\kappa$ \textbf{then return} $L_x[y]$;
 $(*$ precomputed result $*)$\\
\>$p:=\Tffloor{{{(y-y_x)}/\kappa}}$;\\
\>$\hax:=\Tfloor{{x/p}}p$;
 $(*$ largest multiple of $p$ no larger than $x$ $*)$\\
\>\textbf{if} $\hax>0$ and
 $\hax\bmod 2 p=0$ \textbf{then} $\hax:=\hax-p$;
 $(*$ if $\hax>0$, ensure $\pi(\hax)=p$ $*)$\\
\>\textbf{return} $L_{\Tvnn{Jump}[\mskip 1mu\haxx\mskip 3mu]}[y]$;
 $(*$ appropriate entry in ladder at $\Tvn{Jump}[\hax]$ $*)$
\end{tabbing}
\caption{The execution of a query $(x,y)$
in the one-level FL structure.}
\label{code:query}
\end{figure}

\begin{figure}
\begin{center}
\includegraphics{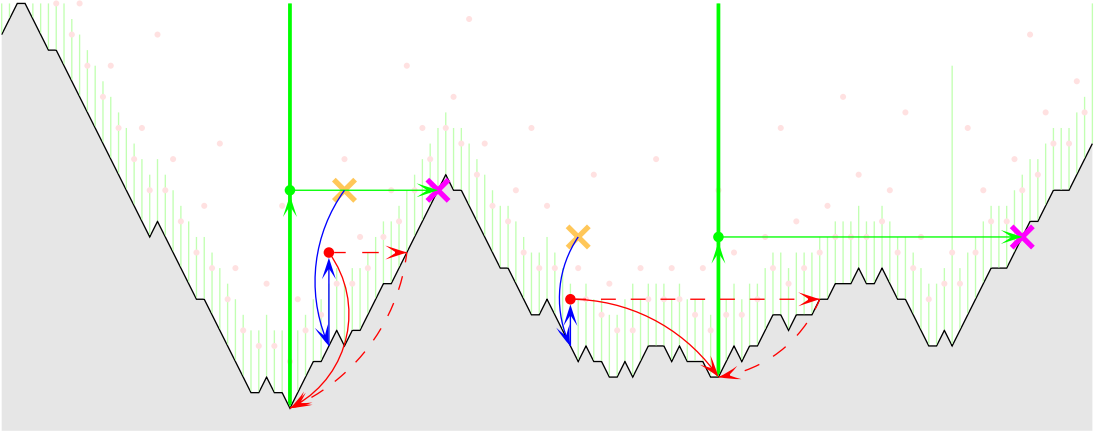}
\end{center}
\caption{The execution of two example queries in the
one-level FL structure with $\kappa=5$.}
\label{fig:hagerup}
\end{figure}

\begin{figure}
\begin{center}
\includegraphics{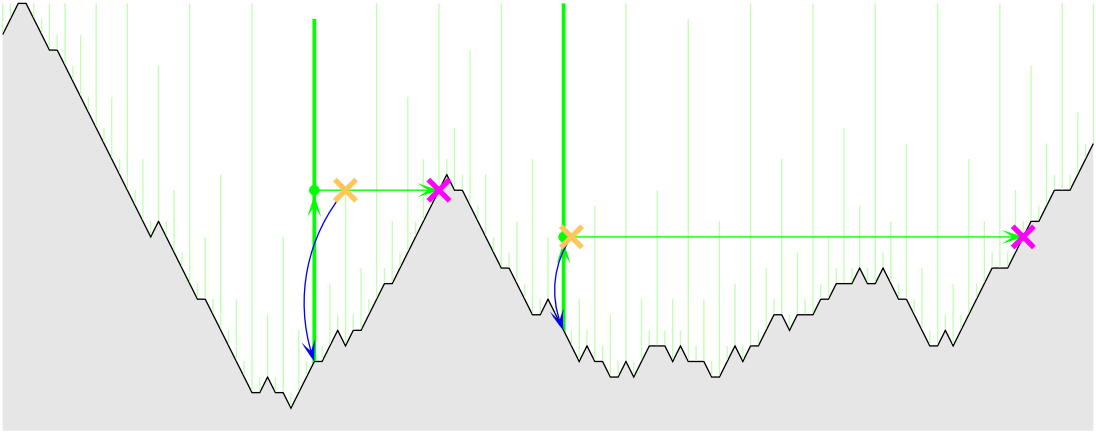}
\end{center}
\caption{The execution of the example queries in Ben-Amram's
data structure.}
\label{fig:ben-amram}
\end{figure}

\begin{figure}
\begin{center}
\includegraphics{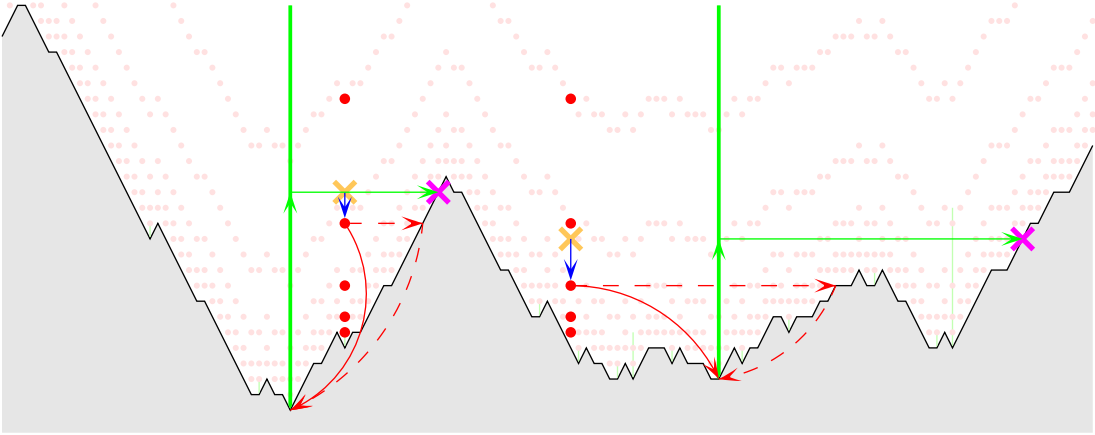}
\end{center}
\caption{The execution of the example queries in the
data structure of Bender and Farach-Colton.}
\label{fig:bender}
\end{figure}

The execution of two example queries in
the one-level FL structure is illustrated
in Fig.~\ref{fig:hagerup}.
An orange cross marks a query $(x,y)$,
and two blue arrows lead from $(x,y)$ first
to $(\hax,y_{\haxx})$
(with $\hax$ as computed in the query
procedure of Fig.~\ref{code:query}) and then to
$(\hax,y_{\haxx}+(\kappa-2)p)$, which
is marked with a red dot.
A red arrow leads from there to
the ``foot'' $(\overline{x},y_{\overline{x}})$
of the ladder at $\overline{x}=\Tvn{Jump}[\hax]$.
The red arrow ``shunts out'' the value
$x'=\Tvn{FL}(\hax,y_{\haxx}+(\kappa-2)p)$,
which is hinted at with dashed red arrows
that pass via $(x',y_{x'})$.
The ladder at $\overline{x}$ is shown in green,
and the entry consulted in the ladder and the
information provided by the ladder are
symbolized by two green arrows.
A magenta cross, finally, marks the point
$(\widetilde{x},y_{\tixx})$,
where $\widetilde{x}=\Tvn{FL}(x,y)$.
The ladders not used by the example
queries are hinted at in pale green.
Figs.\ \ref{fig:ben-amram} and~\ref{fig:bender}
show, using similar drawing conventions,
how the same queries are executed in data structures
derived from the basic data structures
of Ben-Amram~\cite{Ben09}
and Bender and Farach-Colton~\cite{BenF04}
by translating them to our setting and
streamlining them where possible.
The ladders of~\cite{Ben09} are of total height
$\Theta(n\log n)$, and the jump tables
of~\cite{BenF04} (shown as columns
of red or pale red dots in Fig.~\ref{fig:bender})
hold a total of $\Theta(n\log n)$ entries,
which explains why these earlier data structures
are less efficient.

\subsection{Correctness}

Assume that the procedure of
Fig.~\ref{code:query} is carried out for
a nontrivial query $(x,y)$ and define
$p$ and $\hax$ as in the procedure.
Let $t=y-y_x$ and
observe that
$\kappa p\le t\le 2\kappa p-1$ and that
$|x-\hax|\le 2 p-1$.
Since $y-y_x=t>|x-\hax|$,
it is clear that no integer $i$ with $\hax\le i\le x$
can have $y_i\ge y$, so
$\Tvn{FL}(x,y)=\Tvn{FL}(\hax,y)$
(informally, nothing blocks the sight
between $(x,y)$ and $(\hax,y)$).
If $\hax=0$, we have $\Tvn{Jump}[\hax]=0$
(by the special convention regarding $\Tvn{Jump}[0]$),
the nontriviality of the query
shows that $y$ belongs
to the index set $\{y_0+1,\ldots,y\Tsub{max}\}$
of $L_0$, and the procedure correctly
returns $L_0[y]$.
Assume from now on that $\hax>0$, so that
$\pi(\hax)=p$.
We shall need the following bounds on $y-y_{\haxx}$.
\begin{align*}
y-y_{\haxx}=t+(y_x-y_{\haxx})&\ge
t-|x-\hax|\ge\kappa p-(2 p-1)=(\kappa-2)p+1\mbox{\quad and}\\
y-y_{\haxx}=t+(y_x-y_{\haxx})
&\le t+|x-\hax|\le(2\kappa p-1)+(2 p-1)=(2\kappa+2)p-2.
\end{align*}

Define $x'=\Tvn{FL}(\hax,y_{\haxx}+(\kappa-2)p)$ and
$\overline{x}=\Tvn{Jump}[\hax]=\Tvn{Valley}[x']$.
In order to demonstrate that the data structure
operates correctly in the remaining cases,
we must show that $y$ belongs to the index set
of $L_{\overline{x}}$ and that
$\Tvn{FL}(x,y)=\Tvn{FL}(\overline{x},y)$.
Assume first that $x'=n\;(=\bot)$.
Then we also have
$\Tvn{FL}(\hax,y)=n$
(because $y\ge y_{\haxx}+(\kappa-2)p$),
$\overline{x}=n-1$
(by the special convention regarding $\Tvn{Valley}[n]$),
$\Tvn{FL}(\overline{x},y)=n$
(because $\hax\le\overline{x}$),
and $y>y_{\overline{x}}$
(because $y>y_i$ for $i=\hax,\ldots,n-1$).
Since $L_{n-1}$ is of height $y\Tsub{max}-y_{n-1}$,
it is clear that $y$ belongs to the index set
of $L_{\overline{x}}$ and that the query
returns the correct value, namely
$\Tvn{FL}(\overline{x},y)=\Tvn{FL}(\hax,y)=\Tvn{FL}(x,y)=n$.
Assume from now on that $x'<n$ and therefore
that $y_{x'}=y_{\haxx}+(\kappa-2)p<y$.

Because $x'=\Tvn{FL}(\hax,y_{x'})$,
$(\hax,y_{\haxx})$ is down-left reachable
from $(x',y_{x'})$.
Since $\overline{x}=\Tvn{Valley}[x']$,
this shows that $y_{\overline{x}}\le y_{\haxx}$.
Another consequence of the relation
$\overline{x}=\Tvn{Valley}[x']$ is that for every
$j\in J=\{y_{\overline{x}},y_{\overline{x}}+1,\ldots,y_{x'}\}$,
the set $I_j=\{i\mid\overline{x}\le i\le x'$ and $y_i=j\}$
is nonempty and $\Tvn{Valley}[\min I_j]=\overline{x}$.
It follows that
$\Tvn{Weight}[\overline{x}]\ge|J|=y_{x'}-y_{\overline{x}}+1$.
Since $\kappa'\ge 1$, we now find
\begin{align*}
y&\le y_{\haxx}+(2\kappa+2)p-2
\le y_{\haxx}+\kappa'(\kappa-2)p-2
=y_{\haxx}+\kappa'(y_{x'}-y_{\haxx})-2\\
&\le y_{\overline{x}}+\kappa'(y_{x'}-y_{\overline{x}})-2
\le y_{\overline{x}}+\kappa'(\Tvn{Weight}[\overline{x}]-1)-2.
\end{align*}
Because $y>y_{\haxx}\ge y_{\overline{x}}$, this shows that
$y$ belongs to the index set of~$L_{\overline{x}}$.
Finally observe that the relation
$y_i\le y_{x'}<y$ holds both for
$\overline{x}\le i\le x'$
(because $\overline{x}=\Tvn{Valley}[x']$)
and for
$\hax\le i\le x'$ (because $x'=\Tvn{FL}(\hax,y_{x'})$).
Thus $y_i<y$ is satisfied for all integers $i$ between
$\overline{x}$ and $\hax$, inclusive, so
$\Tvn{FL}(\overline{x},y)=\Tvn{FL}(\hax,y)=\Tvn{FL}(x,y)$.
Therefore the query returns the correct result
also in this final case.

\subsection{Time and Space Requirements}

The function $x\mapsto\Tffloor{x}$ can be evaluated
in constant time.
E.g., this can be done by using the
\texttt{bsr} instruction supported by modern CPUs or
by lookup
in tables that can be constructed in $O(\sqrt{n})$
time and occupy $O(\sqrt{n})$ words.
Similarly, it is easy to compute
$\pi(x)$ for $x=1,\ldots,n-1$
in average constant time per value
by inspecting the bits in the binary representation
of $x$ in the order from right to left
until a \texttt{1} is encountered.
It is now obvious that the initialization of the
1-level FL structure takes $O(n)$ time and
that it answers every query in constant time.
In addition to a small number of simple variables,
the data structure must store the array \Tvn{Jump},
of $n$ entries, and the ladders $L_0,\ldots,L_{n-1}$.
The two ladders $L_0$ and $L_{n-1}$ are of height
at most $y\Tsub{max}-y\Tsub{min}\le n$ each,
and the ladders $L_1,\ldots,L_{n-2}$ are of total
height at most
$\sum_{x=1}^{n-2}\max\{\kappa-1,\kappa'(\Tvn{Weight}[x]-1)-2\}
\le(\kappa-1+\kappa')n$,
where the inequality follows from the fact that
$\sum_{x=0}^{n-1}\Tvn{Weight}[x]=n$.
The factor $\kappa-1+\kappa'$ is a constant,
for every fixed $\kappa$, that takes
on its minimum value of~8 for $\kappa\in\{4,5\}$.
Thus it is clear that the data structure occupies
$O(n)$ words or $O(n\log n)$ bits.
During its construction $O(n)$ additional words are
needed for the arrays
$S[0\Ttwodots n]$, $\Tvn{Valley}[0\Ttwodots n]$,
$\Tvn{Weight}[0\Ttwodots n-1]$ and
$\Tvn{RightSight}[y\Tsub{min}\Ttwodots y\Tsub{max}+1]$.
It is easy to reduce the space requirements of the
finished data structure by a constant factor at the
price of a somewhat higher (but still constant)
query time.
E.g., for all integers $x$ and $x'$ with
$0\le x\le x'<n$ and all integers $y$,
$\Tvn{FL}(x,y)=\Tvn{FL}(x',y)$ unless
$y_i\ge y$ for some $i\in\{x,\ldots,x'\}$,
so it is possible to do away with the bottom
$k$ entries of every except every $\ell$th ladder
for arbitrary fixed positive integers
$k$ and $\ell$ with $k+\ell\le\kappa$.
Another possibility is to equip each element
of $\{1,\ldots,n-1\}$ with two ``jump values'',
rather than one.
We have reproved the following result of
\cite{BerV94,Die91,AlsH00,BenF04,Ben09}.

\begin{theorem}
\label{lem:basic}
Given a 1-difference sequence $Y$ of length $n$,
a data structure that answers
FL queries about $Y$ in constant time and occupies
$O(n\log n)$ bits can be constructed in $O(n)$ time.
Given an $n$-node rooted tree $T$,
a data structure that answers
LA queries about $T$ in constant time and occupies
$O(n\log n)$ bits can be constructed in $O(n)$ time.
\end{theorem}

\bibliography{arx}

\end{document}